\documentclass[pdflatex,sn-mathphys-num]{sn-jnl}

\usepackage{graphicx}%
\usepackage{multirow}%
\usepackage{amsmath,amssymb,amsfonts}%
\usepackage{amsthm}%
\usepackage{mathrsfs}%
\usepackage[title]{appendix}%
\usepackage{xcolor}%
\usepackage{textcomp}%
\usepackage{manyfoot}%
\usepackage{booktabs}%
\usepackage{algorithm}%
\usepackage{algorithmicx}%
\usepackage{algpseudocode}%
\usepackage{listings}%

\theoremstyle{thmstyleone}%
\theoremstyle{thmstyletwo}%

\theoremstyle{thmstylethree}%

\begin{document}

\title[A new monetary metric in the thermodynamic relation between energy and GDP]{A new monetary metric is found in the thermodynamic relation between energy and GDP.}

\author*[1,2]{\fnm{Brian P.}\sur{Hanley}}\email{brian.hanley@bf-sci.com}

\affil*[1]{\orgname{Butterfly Sciences}, \city{Post Falls}, \state{ID}, \country{USA}}
\affil[2]{\orgname{UTRGV School of Medicine}, \city{Edinburgh}, \state{TX}, \country{USA}}


\abstract{A robust thermodynamic relation between inflation corrected monetary valuation and energy emerges from existing work. This is based on the energy used, the aggregate efficiency of all production processes ($\Lambda(t)$) in terms of Joules per dollar of gross world product, and the gross world product:  $\frac{E_A(t) \text{[J]}}{\Lambda(t) \text{[J/\$]}} = Y(t) [\$]$  where: J = Joules, \$= currency. This directs us to the production system and all of its processes in addition to alternatives to carbon energy.  

The original relation appeared in \emph{'Are there basic physical constraints on future anthropogenic emissions of carbon dioxide?'} (\textit{\citeauthor{Garrett2011Arethe} 2011}). There a foundation assumption was made that a variable $\lambda$ representing energy per dollar would disprove the presented model. However, because $\lambda$ has dimension [$\frac{E}{\$ \; GWP}$], it represents the aggregate efficiency of \emph{all} global production, and cannot be a constant in an economic model. Thus, aggregate production efficiency is: $\Lambda(t) \equiv \sum {\lambda_i(t) \cdot \frac{P_i}{GWP}}$. The claimed 50 year constant relation of $W$ ($\sum_{i = n}^{t} Y_i \text{ [\$]}$) to the energy $E$ of the final year is incorrect---the relation is not flat, nor should this be expected. The graph of $W$ from 1970 back in time is shown to have an historic minimum in 1970 driven by growth in energy consumption and increasing efficiency of energy use that is unlikely to be repeated. 

With improvements, a robust thermodynamic model is obtained that has general application to the relationship between money and energy and may be useable for evaluating the health of currencies and economies.}

\keywords{thermodynamic economics, Cobb-Douglas, Energy Based Cobb-Douglas, Garrett, Lotka's Wheel}



\maketitle

\section{Introduction}\label{sec1}

Between 2011 and 2022, a series of 6 papers appeared based on a constant ratio of energy to cumulative GDP that make thermodynamic arguments regarding economic activity, the relationship of energy and monetary valuation, and the wealth of global society \citep{Garrett2011Arethe,Garrett2012ModesOfGrowthInDynamicSystems,Garrett2012NoWayOut,Garrett2014LongRun1,Garrett2015LongRun2, Garrett2022LotkasWheel}. This work generated significant interest from economists interested in monetary value and money, as well as its relationship to carbon in the economy. From 2010-2025 the foundation paper of the set, \textit{\citeauthor{Garrett2011Arethe} 2011} gathered 56 citations in peer reviewed journals, theses, or scholarly books, at a mean and median rate of 3.5 and 4.0 per year. This foundation article has been cited by economists such as Carsten Herrmann-Pillath (9X), Andrew Jarvis (5X), Felix Ekardt (3X), Robert Ayres (2X), Rafael Lahoz-Beltra (1X), Ugo Bardi(1X), and Steve Keen (1X). From 2022 to 2025, this foundation article has been peer review cited 12 times indicating ongoing interest. There are several current pre-print citations, indicating future citation activity. This signals significant ongoing interest within economics. 

I found this thread of papers of particular interest for several reasons. Primary among them is the effort to put limits on the growth of energy consumption, and hence on carbon fuel consumption. I spearheaded a separate effort to project carbon demand, that had a different basis \citep[p 29, eq. 11]{Hanley2026SocialCostOfGreenhouseGases---OPTiMEMandtheHeatConjecture_s}.  The graph of that OPTiMEM carbon forcing function is reproduced in the supplement §3.51, as figure A. The \textit{\citeauthor{Garrett2011Arethe} 2011} approach is radically different, working from first principles to make a thermodynamic argument, and hence, quite intriguing.

The starting point is the IPCC SRES model (eq. \ref{eq:SRES_1}), which uses an $i$ factor, $\frac{a}{P}$, described as 'energy intensity' per dollar of production (GDP). The most proper SI dimension of $\frac{a}{P}$ is Joules (J) per dollar $\frac{\text{J}}{\$}$. The derivative $\frac{d}{dt}(\frac{a(t) \text{[J]}}{P(t)  \text{[\$]]}})$ has dimension $\frac{watt (W)}{\$}$ because a Joule is a watt second ($W \cdot sec $) and the second disappears as the time interval goes to zero. This 'energy intensity' of $\frac{J}{\$}$ is also the efficiency with which energy is used to create goods and services valued in dollars (or by extension any currency). 

The $\lambda$ introduced in \textit{\citeauthor{Garrett2011Arethe} 2011} also has dimension of joules per dollar $\frac{J}{\$}$, and is related to $i$ and $f$ (eq. \ref{eq:SRES_1}).

\begin{align}
\begin{split}
\label{eq:SRES_1}
    E_m &= p \cdot g \cdot i \cdot c \\
    \text{where: }&\text{each single letter symbol is a function, } \\
    E_m &= \text{CO$_2$ emissions, } p= \text{population, } g=\text{real dollar per capita production }(P), \\
    i&=\frac{1}{f}=\frac{a}{P} \Rightarrow (f=\frac{P}{a}, a= \text{ primary energy, and } \lambda \approx i), \\
    c&=\frac{E_m}{a} \text{ carbonization of energy supply \citep[438]{Garrett2011Arethe} }\\ 
    &\text{(Here I use $E_m$ instead of $E$ to differentiate it from using $E$ for energy.)}
\end{split}    
\end{align}

This article lays out the revised, dimensioned, equations that result from proper understanding of the $\lambda$, and finds that $\lambda$ is a key metric of economies, relating energy to production and money. Thus, it is important to understand the issues identified in the original model that led to this better understanding. One should consider in this context that Sir Isaac Newton made a naive physics-based error about money causing the United Kingdom to pay more for the gold than minted coins were worth, and another giant of economics, Adam Smith, did not seem to understand the problem Sir Isaac created \citep[p. 120-121]{Gardiner2006EvoCreidtaryControls}. 
\begin{enumerate}
    \item The foundation miscue: Based on a theoretical model drawn from growth of droplets and snowflakes, energy per dollar of cumulative GWP was believed to be constant, or the model would be disproved. Dimensional analysis provides the bridge to clarification that this assumption cannot be true. From this assumption two problems appeared. 
    \begin{enumerate}
        \item Data problem: Claim of a constant cumulative GDP ratio to final year's energy in inflation-adjusted dollars \cite{Garrett2011Arethe} (\textit{'9.7 ± 0.3 mW per inflation-adjusted 1990 US dollar'}) found in empirical data since 1970, representing the majority of human energy use. This constant relation is not present in the empirical data, only in a special dataset (Fig. \ref{Fig-1_Y_E_Y-E}-F \& \ref{Fig-1_Y_E_Y-E}-E). 
        \item Mathematical fallacy: Assuming the $\lambda$ variable (energy per dollar ratio) must be constant resulted in calculus that generated mutually conflicting equations and a divide by zero error \citep{Garrett2022LotkasWheel}. Believing $\lambda$ must be constant caused this fallacy to be applied to the real world. Inflation dynamics relative to energy were naively interpreted on the basis of the fallacy to mean that inflation could make real GDP go to zero. 
    \end{enumerate}
    \item An attempt to use Market Exchange Rate (MER) is made instead of Purchasing Power Parity (PPP) for a dataset going back to 1 CE \citep{Garrett2022LotkasWheel}. Early IPCC reports from scientists often used MER, which resulted in diligent attempts to redress this, and the difficulty of doing so is visible in a 2004 Stanford hosted conference \cite{Heston2004TheFlawOfOnePriceSomeImplicationsForMER-PPP_Discussions}. I suggest this transition was simply missed. 
    \item An energy dataset created going back to 1 CE was naive. Year 1 CE energy is not approximately equal to estimated energy use from 1902 CE (Fig. \ref{Fig-1_Y_E_Y-E}-C). 
\end{enumerate} 

With corrections and improvements, the \textit{\citeauthor{Garrett2011Arethe} 2011} concept results in a valuable contribution that may have major impact. It is my view that it represents a flash of insight, and the resulting model showed results tantalizingly close to some empirical data. I found the disbelief that humans in the year 1 CE could have lived with energy consumption below energy at the turn of the 20th century completely sincere. 

The equations resulting from these corrections should be important to economists, as well as useful to climate modelers as \textit{\citeauthor{Garrett2011Arethe} 2011} originally intended. I have endeavored to make this set of equations easy to understand and use. Thus, I took a cue from recent work on energy-based Cobb-Douglas \citep{KEEN2019NoteOnEnergyInProduction} for my function names, and show dimensions. Dimensional analysis can prevent many errors. 

Impact: 
\begin{enumerate}
    \item I think that $\Lambda$ and $\lambda$, which are related to SRES $i$ and $f$ (eq. \ref{eq:SRES_1}), are important metrics that tell us the health and progress of the real economy and a nation's money. Improving monitoring of exergy, drilling down into usage and finding ways to capture and report exergy from industry should have significant payoffs in our quest to get more for less.  
    \item A primary impact is that that we aren't entirely prisoners of the long arm of history, as suggested. The equations and development of $\lambda$ point to where our flexibility is.  
    \item To provide a clean, formalized set of equations that can be used going forward in a variety of ways, including with Energy Based Cobb-Douglas and other, similar efforts. 
    \item To nullify the idea that the gross world product should go to zero in real terms because energy growth goes to zero. 
\end{enumerate} 

\section{Corrections and improvements}
\textit{'Are there basic physical constraints on future anthropogenic emissions of carbon dioxide?'} \citep{Garrett2011Arethe} is the inception paper that lays out the atmospheric physics model of growth of droplets/snowflakes that was adapted to economic growth. This first model uses one set of symbols. The most recent paper is  '\textit{Lotka’s wheel and the long arm of history: how does the distant past determine today’s global rate of energy consumption?}' \citep{Garrett2022LotkasWheel}. This latter paper uses different symbols in some instances, and presents a set of propositions and conjectures.  

\textit{Garrett 2011} thermodynamically models a society as similar to an organism that starts from  a single cell, and as it grows to adulthood, represents the sum of all of the work required for it to grow and maintain itself \citep[pp 438-443]{Garrett2011Arethe}\footnote{ The meaning of the symbol $w$ for work in  “\emph{Are there basic physical constraints on future anthropogenic emissions of carbon dioxide?}' \citep{Garrett2011Arethe} is not the same as the $w$ of \textit{Garrett 2022} \cite{Garrett2022LotkasWheel}. Thermodynamic work in the former is the physics definition (See supplement). In this economics use, production which is technically the results of work, is used as a surrogate. Production is measured in dollars as GDP.}. This concept is projected into economics as the summation of all GWP throughout history, founded in the idea that current society is built on everything that came before it. 

This summation function of all historical production is named \textit{C(t)} in \textit{Garrett 2011}\citep[p. 441-2. Eq's. 4-7]{Garrett2011Arethe}, and is termed \textit{W(t)} in \textit{Garrett 2022} \citep[p. 1022 Eq. 1]{Garrett2022LotkasWheel}. 

\textit{Garrett 2022} uses GDP as GDP for the globe, which can be confusing. Here I use gross world product ($GWP$). $Y$ is the symbol used in equations of \textit{Garrett 2022} for $GWP$ by year, which conforms to a common economics use to represent production, so I use $Y$ as much as possible instead of $P$. I also make use of replicate datasets as part of the validation method, and use subscripts for dataset symbols to identify them. 
\subsection{The key matter: $\lambda$, the industrial production symbol, cannot be a constant---it is an aggregate that will vary, and improve with energy efficiency }
\label{Sect:Lambda_cannot_be_constant}
This fundamental issue appears because the model used as a template for the economic model \citep[appendix A]{Garrett2011Arethe}, is borrowed from growth of a snowflake or droplet. This atmospheric science model is also discussed using a human child growing up metaphor \citep[p. 440]{Garrett2011Arethe}. In both cases, snowflakes and human children grow in equivalent units by a mechanism that can be considered a singular, unchanging mechanism. This foundation fact of atmospheric physics led to the incorrect belief, “\emph{If $\lambda$
is not constant with time, then the thermodynamic framework is false.}' \citep[pg 443]{Garrett2011Arethe}. From this follows the necessity to find confirmation that $\lambda$ is a constant, and hence, that there is a flat-line relationship to be found between energy and some variant of GWP expressed in monetary units. 

Translated to SI units, $\lambda$ is defined with dimension $\frac{J}{\$}$ \citep[Pg 441, eq.(4)]{Garrett2011Arethe}. That $\lambda$ has dimension of Joules per dollar of GWP is a strong clue to the correct answer.  It means that $\lambda$ measures how energy is used to convert resources into utility economy products, which products are each priced using currency, and reported as GWP. A bit of thought makes clear that $\lambda$ is a measure of the efficiency of industrial production machinery/processes ($K$, the means of production). 

The limits of $\lambda$ are determined by limits on efficiency of conversion of energy into useful work in the service of production. 

Drilling down into GWP, this step of conversion of materials into products (industrial production) using energy incorporates a huge population of different industrial processes with widely varying efficiency of conversion, in upwards of 359 million companies globally \citep{Dyvik2024_359MillionCompanies}. One company can have hundreds of processes, each with different efficiency. These processes change yearly, and even day by day, as do the products produced. Entirely new products having value appear, and old products disappear. Machinery for production is improved and replaced, or machinery with entirely new capabilities installed. As an example, in a factory I automated, an average of over 100 changes to the automobiles being manufactured were implemented per day. This in turn drove installation and upgrades to equipment. 

$\lambda$  represents a large aggregation of equation outputs varying in $t$ describing production efficiency of firms relative to energy.  This aggregate will have new functions introduced and old ones removed, and the functions for each process can change rapidly. \emph{Consequently $\lambda$ cannot be a constant.} Thus, I represent the aggregate of industrial production (eq. \ref{eq:Lambda}) with a capital $\Lambda$. This can also drill down into each firm, as most firms contain many specific industrial processes that use energy and a similar equation will apply that sums all those processes to produce a single $\lambda_i$ for the firm. 
\begin{align}
\begin{split} 
	\qquad\Lambda(t)\;[\frac{J}{\$}]  &\equiv \sum {\lambda_i(t)\;[\frac{J}{\$}] \cdot \frac{P_i \;[\$]}{GWP \;[\$]}}\;\\
    \text{Where: } \lambda_i &=\text{ production efficiency for one of 359 million+ firms globally. } [\frac{J}{\$}] \\
    P_i &= \text{production of a firm [\$]} \quad GWP = \text{ gross world product [\$]}
    \label{eq:Lambda}
\end{split}
\end{align}  
  
Thus, the global aggregate efficiency of converting energy to dollars of product ($\Lambda$) is the sum of the efficiencies of each process used where each process effiency ($\lambda_i$) is weighted by the production's ($P_i$) fraction of gross world product (GWP). 

\subsection{Revised system of foundation equations}

All energy functions are represented with an $E$ and a subscript. Dimensions are shown in brackets,
 J = Joules, \$ = US dollars. 
 
$E_G(t)$ [J] = Gibbs free energy yield from consumption of some fuel source.  Originally $\equiv$ $\Delta G=\Delta H- T \Delta S$ where $\Delta H$ = enthalpy, T = temperature, S = entropy, which symbol is reserved to mean the energy theoretically available.  \\

$E_A(t)$ [J]  = $\alpha$(t) $\cdot$ $E_G(t) \; \text{[J]} $ The available captured energy. (Ex. electrical power generated.)  Originally $\equiv$ $a$

$\alpha(t)$ = $\frac{E_A(t) \;\text{[J]}}{E_G(t) \;\text{[J]}}$ where $0 < \alpha(t) < 1$. The efficiency of conversion of $E_G(t)$ to $E_A(t)$.  Originally $\equiv$ $\alpha$, a constant. 

$E_X(t)$ [J] = $\varepsilon(t)$ $\cdot$ $E_A(t)$ [J] The net exergy.  Originally $\equiv$ $w$ [J]. (This $w$ is not the $w$ of \textit{\citeauthor{Garrett2022LotkasWheel} 2022}.)

$\varepsilon(t)$ = $\frac{E_X(t)\text{ [J]}}{E_A(t) \text{ [J]}}$  where $0 < \varepsilon(t) < 1$. The efficiency of conversion of $E_A$ to useful work $E_X$. Originally $\equiv$ $\varepsilon$ a constant. 

See appendix \ref{Appendix:alpha_EfficiencyOfEngines} for discussion of $\alpha$ and $\varepsilon$ as time variant. See supplement for discussion of equations 2-11. \\

$\frac{E_A(t) \;\text{[J]}}{\Lambda(t) \;\text{[J/\$]}}$ is production.  In \textit{\citeauthor{Garrett2022LotkasWheel} 2022} this is represented by $Y(t)$. 

Thus, $Y(t)$ [\$] $\equiv$  $\frac{E_A(t) \;\text{[J]}}{\Lambda(t) \;\text{[J/\$]}}$

Putting this together, the basic system of equations is: 
\label{Sect:System_of_Equations}
\begin{align}
	Y(t) [\$] &= \frac{E_A(t) \; \text{[J]} }{\Lambda(t) \; \text{[J/\$]}} \\     
	\qquad \Lambda(t) [\frac{J}{\$}] &= \frac{E_A(t) \; \text{[J]} }{ Y(t) \; \text{[\$]} }   \label{eq:Lambda_t}\\
	\qquad E_A(t) [J] &= Y(t) [\$] \cdot \Lambda(t) [J/\$]  \\
   \qquad  \frac{Y(t) \;\text{[\$]}}{ E_A(t) \; \text{[J]}} &= \frac{1}{\Lambda(t) \; \text{[J/\$]}}  \label{eq:1_over_Lambda}
\end{align}
 \qquad

See Appendix \ref{Appendix:CobbDouglas} for the relationship of these equations to a revision of the Cobb-Douglas equation. See supplement for a more complete discussion of equations. 

\subsection{$W(t)$: historical $GWP$ summation equations with notations}
\label{sect:W(t)_Summations_Y_and_pop}
$W(t)$ is the sum of all $GWP$ dollars in the past to the present as defined in eq. \ref{eq:W(t)}.
\begin{align}
\begin{split}
 W(t) \text{ [\$]} &= \sum_{i = n}^{t} Y_i \text{ [\$]}  \Rightarrow   W ( t ) \text{ [\$]} = \int_n^{ t } Y ( t ) \text{ [\$]} dt \qquad \text{\citep[eq. 1]{Garrett2022LotkasWheel}}
\label{eq:W(t)} \\
 \text{Where: } W(t) &= \text{ historical sum of all previous } Y_i \\ 
 Y_i \, &= \text{GWP for a year }i, \quad t = \text{ time in years}   
\end{split}
\end{align} 

If $Y_i$ = 0 this means that humans ceased to exist, because if humans exist, then some $Y_i$, however small, will exist. If humans do not exist we have reached an end, which is not interesting.  So, assume $Y_i > 0$. 

Per equation \ref{eq:W(t)}, if  $Y_i > 0$, $W(t)$ must increase each year by a positive $Y_i$. 

$Y(t)$ is related to population by changing multipliers per Morris \cite[pp. 55, 61, 68, 111]{Morris2013TheMeasureOfCivilization}. 

$Y(t)$ relates to $W(t)$ as $W(t) \; –\; W(t-1) $. $\frac{dW(t)}{dt}$ = $Y(t)$. 

Population ($Pop$) is a driver of $Y(t)$. $\frac{dPop}{dt}$ tracks closely in practice to $\frac{dW}{dt}$ (not shown) as discussed by Garrett et al. \cite[p 1027]{Garrett2022LotkasWheel}.

\begin{figure}[hbt!]
\includegraphics[width=1.0\textwidth]{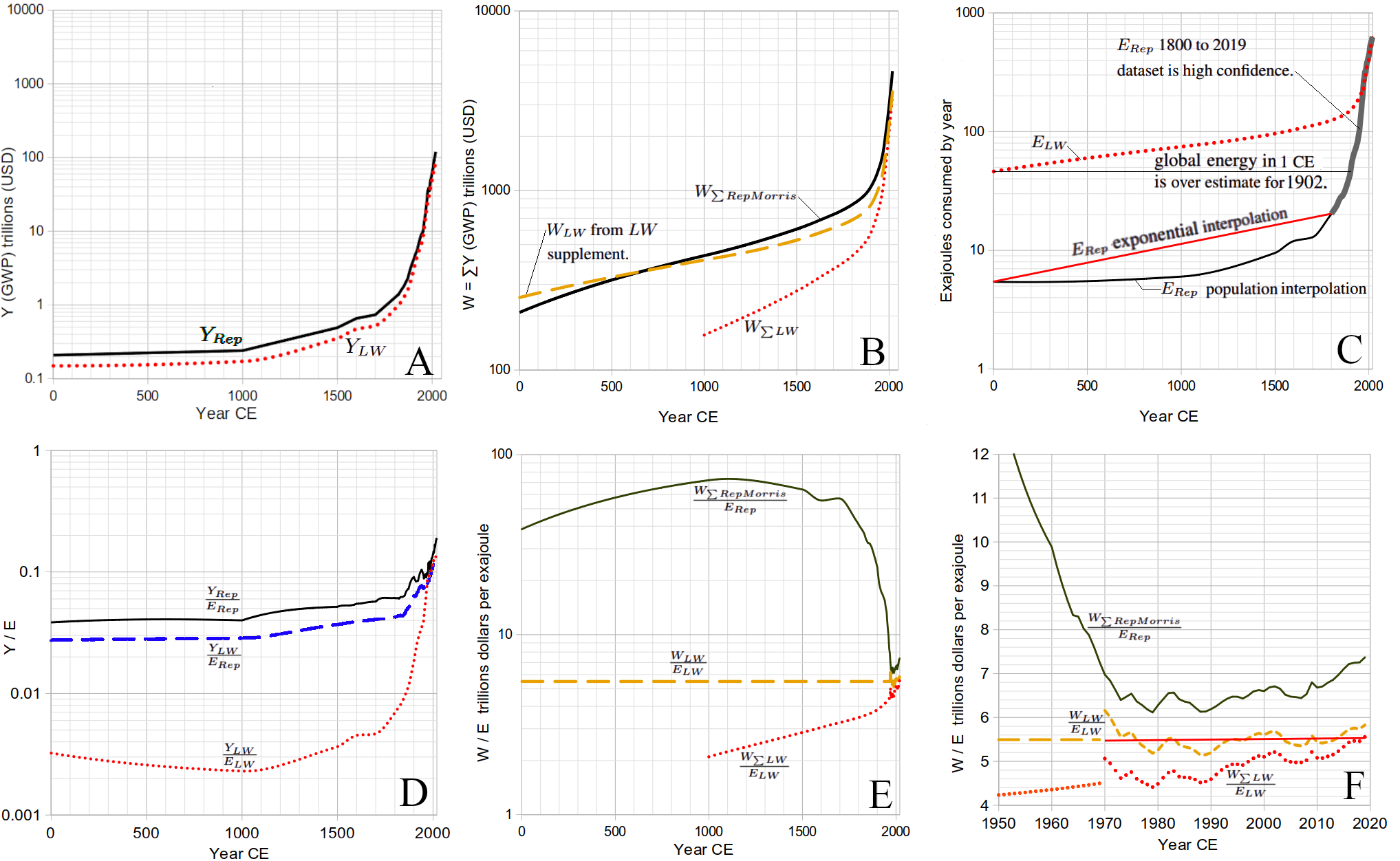}
\caption{ Here dotted red curves are either \textit{Garrett 2022} ($LW$) datasets, or created from them. Dashed gold curves are special datasets also from Lotka's Wheel supplement. Solid black curves are the author's replicates. \textbf{Panel A} Dotted red curve is $Y_{LW}$ (GWP). Solid black curve $Y_{Rep}$ is the interpolated composite. $Y_{Rep}$ is reasonably close, so this is not a primary reason for a large discrepancy. \textbf{Panel B}. $W$ summation curves. Red dotted curve is $W_{\sum LW}$ produced by summation of $Y_{LW}$ dataset, with the first 1000 years truncated due to unrealistic zero start. Black solid curve $W_{\sum RepMorris}$ includes $Y$ estimates to -14,000 CE. Gold dashed $W_{LW}$ curve is supplied directly from $LW$ supplement. \textbf{Panel C}. Exajoules consumed by year. Solid heavy curve is $E_{Rep}$ from 1800 to 2019, which dataset is high confidence. From the 1800 point of roughly 20 exajoules, $E_{Rep}$ red interpolation curve is exponential (eq. \ref{eq:N_i_eq_N_1.a^x}), and black solid curve is population based interpolation to year 1, anchored in Morris \citep{Morris2013TheMeasureOfCivilization}. The population interpolation for $E_{Rep}$ was used going forward. The dotted red curve $E_{LW}$ is anchored in an overestimated value from \textit{Garrett 2022}. \textbf{Panel D}. This shows the effect of the population-based $E$ interpolation on the $\frac{Y}{E}$ ratios, which goes quite flat. The black solid curve $\frac{Y_{Rep}}{E_{Rep}}$ and blue dashed curves $\frac{Y_{LW}}{E_{Rep}}$ are close enough to show that whether the $\pi$ adjustment for $Y_{LW}$ is made or not, it is $E$ that matters. \textbf{Panel E}. $\frac{W}{E}$ ($W$ summations of $Y$, over E).  Black solid $\frac{W_{\sum RepMorris}}{E_{Rep}}$ curve, is summed back to -14,000 CE. The $\frac{W_{LW}}{E_{LW}}$ gold dashed curve uses the $W$ dataset directly from the supplement, with reasonable fit to $\frac{W_{\sum RepMorris}}{E_{Rep}}$. The first 1000 years of $\frac{W_{\sum LW}}{E_{LW}}$ are truncated as in panel B. \textbf{Panel F} $\frac{W}{E}$ detail of panel E from 1950-2019. Black solid curve $\frac{W_{\sum RepMorris}}{E_{Rep}}$. The dotted red curve is $\frac{W_{\sum LW}}{E_{LW}}$ from 1970-2019. The dashed gold curve shows $\frac{W_{LW}}{E_{LW}}$ from $Lotka's Wheel$ supplement. Red horizontal line is a linear fit to the $\frac{W_{LW}}{E_{LW}}$ data from  1970-2020, with equation $f(x) = 1.18E-3x + 3.131$. }
\label{Fig-1_Y_E_Y-E}
\end{figure}

See fig. \ref{Fig-1_Y_E_Y-E}-A \& B and fig. \ref{Fig_Morris_W_E_W-E_Pop}-A, B \& F.

 \subsection{The constant monetary ratio of $w =\frac { W \text{ [\$]}}{ E \text{ [J]} }$ is falsified.} 
\begin{figure}[t!]
\includegraphics[width=1.0\textwidth]{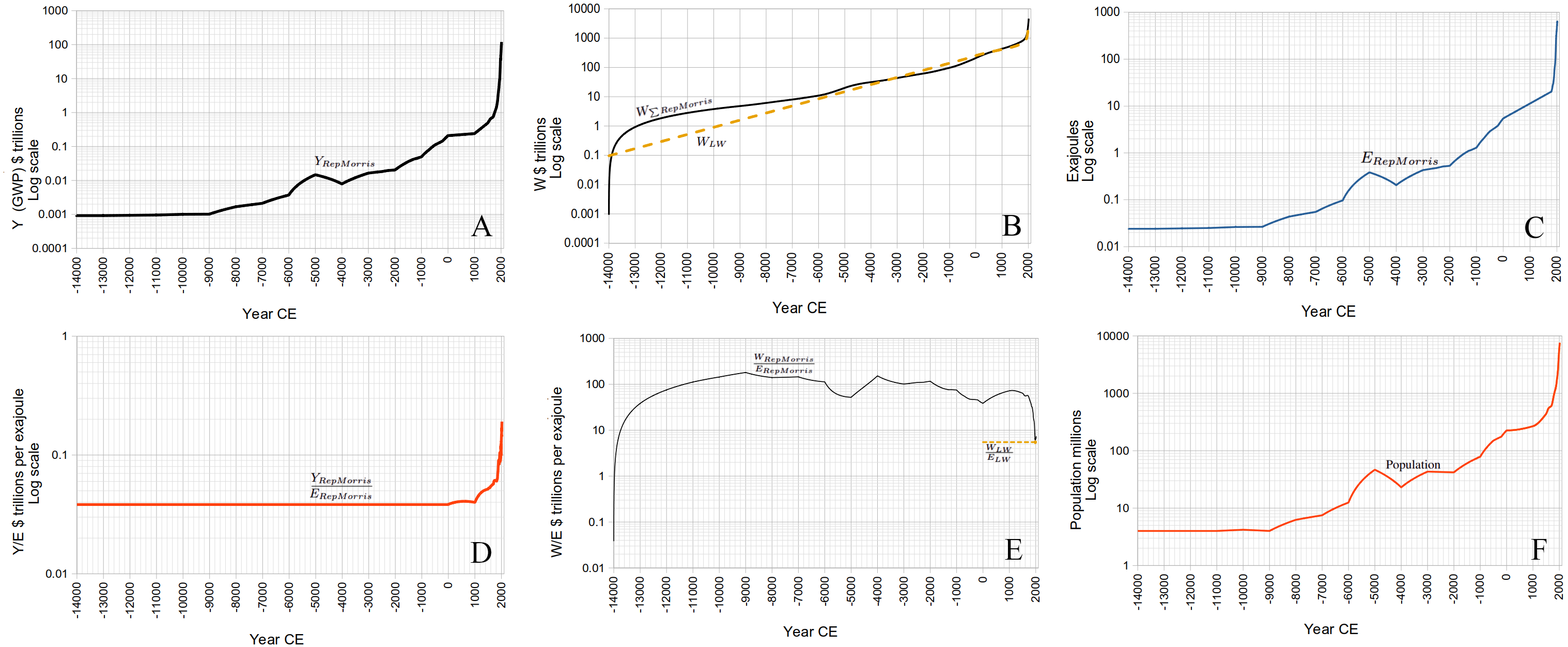}
\caption{Curves extended back to -14,000 CE using Morris tables \citep[pp. 55, 61, 68, 111,]{Morris2013TheMeasureOfCivilization}. Population is always the overwhelming influence. \textbf{Panel A.} $Y_{RepMorris}$ curve is estimated GWP from -14,000 to 2019 CE. \textbf{Panel B.} $W$ curves. $W_{RepMorris}$ is the summation from -14,000 to 2019 CE utilizing the Geary-Khamis dollar kCal method pegged to year 1 CE. $W_{LW}$ is the supplement dataset of \textit{Garrett 2022} from 1 CE forward. A curve fitted to the $W_{LW}$ dataset from 1 CE to 1969 CE was used to project back to -14,000. $W = 244.064 \cdot e^{(5.596E-4) \cdot x}$ ($R^2=0.943$) These fitted curve results are presumed to implement the intent of \textit{Garrett 2022}. By inspection there is fair correspondence between this extended theory-driven curve and the Morris based estimates over an 8,000 year period. The estimate curve does not fit after 1969. \textbf{Panel C}. $E_{RepMorris}$ exajoule estimates based on Morris' kCal figures. \textbf{Panel D.} $\frac{Y_{RepMorris}}{E_{RepMorris}}$ by inspection this curve is quite flat from 1 CE to 1,000 CE. Here, the dependence of Morris based estimates on population is visible in deep history when economies were not monetized for most of it. The Geary-Khamis concept used to develop both $E$ and $Y$ estimates is that grain/feed kCal, which is then extended to total kCal, regulates the pre-history economy. Total energy bottoms out at $\approx$4,000 kCal per capita per day.  These rough estimates are from a time when total energy available did not vary a great deal, and energy efficiency was poor. However energy of rivers and sail is not accounted for here. Improving these will wait for archaeologists, but it is difficult to justify very large deviations from the figures graphed here.   \textbf{Panel E.} Visible by inspection in panel E is that the hypothesis that $\frac{W}{E}$ is a constant is falsified. It also appears that the period from 1970 to 2019 is not part of a long-term trend. Solid black curve is $\frac{W_{RepMorris}}{E_{RepMorris}}$. The short horizontal dashed gold curve is $\frac{W_{LW}}{E_{LW}}$. \textbf{Panel F.} Population estimates. Note that if Morris data from 1 CE to 1800 CE were made use of, it is likely that the interpolation of $E$ in panel B would track closer to population during this period. Morris tables provide GDP/capita/day estimates. Consequently population has an overwhelming influence on $E$ until the industrial revolution.}
\label{Fig_Morris_W_E_W-E_Pop}
10\end{figure}

 The novel measure $\frac { W \text{ [\$]}}{ E \text{ [J]} }$ \cite[eq. 2]{Garrett2022LotkasWheel} (fig. \ref{Fig-1_Y_E_Y-E}-E \& \ref{Fig_Morris_W_E_W-E_Pop}-E) is related to the reciprocal of $\Lambda(t)$, $\frac{1}{\Lambda(t) \; \text{[J/\$]}}$. However, $\frac { W \text{ [\$]}}{ E \text{ [J]} }$ is quite different, because the numerator is the all-history summation, but the denominator is energy consumed in the final year. This novel measure is intended to be an indicator of how sustainable a civilization is, what the limits to growth are. Thus, if the underlying model had a growth constant, that constant should be visible. 
 
 Empirically, in figure \ref{Fig-1_Y_E_Y-E}-E \& F, and \ref{Fig_Morris_W_E_W-E_Pop}-E, $\frac{W_{LW}}{E_{LW}}$ is the  curve from the $W/E$ column in the \textit{\citeauthor{Garrett2022LotkasWheel} 2022} spreadsheet supplement. $\frac{W_{\sum LW}}{E_{LW}}$ is the curve with numerator calculated from the $Y$ column of this supplement. $\frac{W_{\sum RepMorris}}{E_{RepMorris}}$ (fig. \ref{Fig-1_Y_E_Y-E}-B) is the curve with numerator calculated using the extended replicate $Y$ and $E$ datasets derived from Morris \citep[pp. 55, 61, 68, 111]{Morris2013TheMeasureOfCivilization}.
 
 The ${w}$ of \textit{\citeauthor{Garrett2022LotkasWheel} 2022}, which is supposed to represent ${\frac{W}{E}}$ (fig. \ref{Fig-1_Y_E_Y-E}-F) is seen by inspection to have an increasing trend for both the ${\frac{W_{\sum LW}}{E_{LW}}}$ and ${\frac{W_{\sum RepMorris}}{E_{Rep}}}$ curves from 1970 forward. After 1970, the ${\frac{W_{\sum RepMorris}}{E_{Rep}}}$ curve is a bit steeper than the ${\frac{W_{\sum LW}}{E_{LW}}}$ curve. By contrast, the $\frac{W_{LW}}{E_{LW}}$ supplement dataset appears flat, having a fitted equation after 1970 with a slope of 1.18E-3, and supports the constant contention.  

Prior to 1970 CE, inspection of the entire ${\frac{W_{\sum RepMorris}}{E_{RepMorris}}}$ graph also falsifies the constant $w$ hypothesis, showing that 1970 CE appears to be a transient minimum (fig. \ref{Fig_Morris_W_E_W-E_Pop}-E) that is unlikely to be repeated. 
 
Mathematically, the ${w}$ of eq. \ref{eq:W_over_E_eq_5.50} has the same dimension as $\frac{1}{\Lambda(t) \; \text{[J/\$]}}$ (eq. \ref{eq:1_over_Lambda}). Because $\Lambda(t)$ is not a constant, $\Lambda(t)$ is collated into $W$, $W$ is the numerator of $w$, the $w$ constant proposition is falsified on this ground. 
  
\begin{align}
\begin{split}
{w} &= { \frac { W \text{ [\$]}} { E \text{ [J]} } }  \qquad \qquad \qquad \qquad  \qquad \qquad  \qquad \qquad \text{\citep[eq. 2]{Garrett2022LotkasWheel}}\\
 \text{Where Units: } E &= \text{ exajoules } \quad W = \text{ currency ( USD ) }\\
 {w} & \text{ dimension} = \frac{ \text{currency ( USD )}}{ \text{exajoule}}  
\end{split}
\label{eq:W_over_E_eq_5.50}
 \end{align}

\subsection{Real economic production approaches zero by inflation: assuming a constant creates a mathematical fallacy.}
	Based on equations \ref{eq:Y_approx_w_x_dDdt} and \ref{eq:Y_estimate_w-hat_eq_Y_over_dE_dt}, the proposition is made that if $\frac{dE}{dt}$ = 0, then $Y$ goes to zero based on eq. \ref{eq:Y_approx_w_x_dDdt}. But this creates a paradox.
 By eq. \ref{eq:Y_approx_w_x_dDdt}, if $\frac{dE}{dt}$ = 0, then $Y=0$. But at the same time, by eq. \ref{eq:Y_estimate_w-hat_eq_Y_over_dE_dt}, ${w} = \frac{Y}{0}$. And ${w}$ is a fixed slope of a line. The only meaning this can have is that ${w}$ has a vertical asymptote that cannot be reached with discontinuity at zero and is a fixed (near horizontal) line at the same time. As the $w$ factor cannot simultaneously change in slope with a discontinuity (eq. \ref{eq:Y_estimate_w-hat_eq_Y_over_dE_dt}) and remain a constant, this is a mathematical fallacy. 
  
\begin{equation}
     Y = \frac { dW } { dt }   \Rightarrow Y = {w}{ \frac { dE } { dt } } \qquad  dE/dt \rightarrow 0  \qquad Y = 0 \qquad \quad \text{\citep[eq. 3]{Garrett2022LotkasWheel}}
\label{eq:Y_approx_w_x_dDdt}
`\end{equation}

\begin{equation}
     {w}={ \frac { Y } { dE/dt }} \qquad  dE/dt \rightarrow 0  \qquad  w = \frac{Y}{0}=\infty \qquad \qquad \qquad \qquad \text{\citep[eq. 4]{Garrett2022LotkasWheel}}
\label{eq:Y_estimate_w-hat_eq_Y_over_dE_dt}
\end{equation} 
\text{ }

The proposition that real-world production disappears caused by inflation/hyperinflation is also falsified on empirical grounds in Appendix \ref{Sect:CPI_dE-dt}. 

The issue presented here disappears if a more appropriate equation, such as eq. \ref{eq:Growth_Yt_Y0xgt} is used. See Appendix \ref{Sect:Growth_equation}. Again, this stems from needing to find a constant for $\lambda$, which is consistent.

\subsection{Market exchange rate (MER) versus purchasing power parity (PPP).}
\begin{quote}
    \textit{Countries with a low standard of living tend to have a relatively high gross domestic product when expressed in PPP rather than market exchange rate dollars because equivalent products and services tend to be less expensive. However, because the focus of this study is energy production and associated CO$_2$ emissions, rather than national standard of living, it is historical records of market exchange rate valuations that are used.} - \cite[p 453]{Garrett2011Arethe}
\end{quote}
Use of MER conflicts with the IPCC \citep{IPCC_2007_ReportTheUseOfMER} recommendation to use PPP. Purchasing power parity was created to normalize valuation of currencies so as to allow comparison across geography and time. The data sources use PPP methods.  Thus, the $\pi$ adjustment created becomes an arbitrary multiplier projecting back in time\footnote{The Greek letter $\pi$ is used for multiple purposes in economics which sometimes causes confusion to non-economists.}. When I examined it, there is no meaningful qualitative difference, as this $\pi$ factor is a constant. Even putting aside the issue that MER will show distortions, real MER adjustment would be highly complex, potentially day by day, and usually fluctuating between nations unless held in place by fiat. The intent of $\pi$ may be laudable, but it is not necessary nor useful.

MER for dollars do not exist for most of the history of the past 2020 years that this $\pi$ factor is applied, so performing this based on records is impossible. The issue raised regarding standard of living is one that PPP estimators are aware of---it is a primary reason for inventing PPP. The US dollar maintains a high value due to its use as international settlement currency, perception of safe haven, and most nations working to maintain as much of a trade surplus with the USA as possible. The USA's trade deficit is legendary, and has not been near zero since 1975 \cite{FRED2025XTNTVA01USA667S}.

\emph{“Purchasing power parity data provide a common measuring rod that allows comparison, not only of India and America now, but of India now with Britain before the industrial revolution,”} \citep{Deaton2010UnderstandingPPPs}.  The operative word here is “allows”. Without PPP such historical comparisons cannot be done meaningfully. The US dollar was created after 1792 by the new United States \citep{Nussbaum1937TheLawOfTheDollar}, and prior to this, there cannot be MER data for it. 

There is mention of Geary-Khamis in a discussion of Maddison data \cite[p 453]{Garrett2011Arethe}. The PPP concept has been embraced by archaeologists, with the Geary-Khamis kCal grain-dollar standard, deriving from van Zanden and others \citep[pp. 67-71]{Morris2013TheMeasureOfCivilization}, that relates food to other forms of energy. Archaeologists may measure early civilizations in terms of human and animal kCal inputs for both energy and valuation of wealth, which is a PPP conceptual approach---used to develop an estimate of deep time production.  

\section{Discussion}
As work progressed utilizing Morris' tables \cite[pp. 55, 61, 68, 111]{Morris2013TheMeasureOfCivilization}, the $\frac{W}{E}$ curves (fig. \ref{Fig-1_Y_E_Y-E}-E \& \ref{Fig_Morris_W_E_W-E_Pop}-E), began to show that the hypothesis that the period from 1970 to 2019 is representative of the past ratios to energy did not hold. The $W_{LW}$ dataset, that was created as an attempted estimate back into the past, tracks to $W_{\sum RepMorris}$ in broad brush, with two possible caveats. First, from 1970 CE forward, the slope is not a fit. Second, deep time production tapers back very far, with stone tools appearing 3.3 mya \cite{Harmand20153.3-million-year-oldStoneToolsFromLomekwi3}, probable genus homo Acheulean tools 1.76 mya \cite{Lepre2011AnEarlierOriginForTheAcheulian}, bone tools 1.5 mya and more sophisticated tools 300-400 kya \cite{delaTorre2025SystematicBoneToolProductionAt1.5millionYearsAgo}. Declaring a starting point of civilization is complicated. 

Second, the modern steep $\frac{W}{E}$ drop from $\approx$1700 CE to 1970, then reversal, is seen to be a local minimum at 1970. (See: fig. \ref{Fig-1_Y_E_Y-E}-E \& F and fig. \ref{Fig_Morris_W_E_W-E_Pop}-E). After 1970 $GDP$  growth finally began to outpace exponential growth in $E$, together with increasing efficiency of $\alpha(t)$ and $\varepsilon(t)$ such that this is registered in the $W$ summation of $Y$.  It makes me uneasy that the relationship to entropy (decay of capital, $K$) may be invisible here, although $Y$ is increasing relative to $E$. My concern is that too much of $Y$ could be churn---GWP with little added utility value. (i.e. Useful life span of production is too short.) Thus, relative to capital ($K$) and its decay (entropy), some of the faster increase in $Y$ over the increase in $E$ may be a kind of fiction, because of churn in the form of value engineering and planned as well as opportunistic \citep{Cooper2024AppleLawsuitiPhones} obsolescence. However, one could also argue that a society that correctly forecasts the social need for $K$ could be more efficient. To what extent this is true could come out in a careful accounting of $K \, [\$/J]$, as well as $L \, [\$/J]$ (labor). This means drilling down into industrial process use of energy to better estimate $E_X(t)$ to raise the productivity.

There needs to be agreement regarding what, exactly, $K$ should include for our purposes---for instance does $K$ include commodities from napkins to plastic forks, or is $K$ just the traditional means of production? There is also the question of what the services economy means relative to $K$. Additionally, this $K$ question gets into the realm of social capital as defined by culture and subcultures (i.e. corporations, ethnic groups), education, institutional knowledge, entrepreneurial spirit, social pathways of contribution to the economy, 'traditional' social capital (one representation of which is the corruption perception index \citep{Lambsdorff1999TransparencyInternationalCorruptionPerceptionsIndex} ) and other “soft capital”. All of these need to be accounted for if feasible, and estimated in deep time, which requires that a common definition for capital specific to the thermodynamic approach ($K_{Thermo}$) needs to be forged. This definition needs to be something that can be grasped and estimated by archaeologists and categorized for those that work on this. Here I present a straightforward method based on Morris \cite{Morris2013TheMeasureOfCivilization} (fig. \ref{Fig_Morris_W_E_W-E_Pop}), however, I do not believe it is necessarily complete. 

I did not pursue any empirical work on $K$, which is less tractable than $GWP$ and energy consumption. When considering $K$, it is intuitively obvious that if physical capital, for instance roads, harbors, airports, spaceports, trains, dwellings, water pipelines and energy installations, etceteras last longer, this can be higher overall efficiency (dollars of production per joule) than our current methods---barring revolutionary replacements. Currently, many civil engineering projects, and certainly homes, are engineered for an assumed life span that is far shorter than the centuries that history would suggest. For such civil engineering projects, value-engineering is not necessarily the same as the planned obsolescence that many manufactured products have engineered into them, even though it effectively generates the same net result of requiring more “churn” of $K$. The “churn” motive of our economic system is problematic. It can signal mature industry that has few or no more innovations to offer that are valuable, or one with a degree of monopoly power that has decided to play rentier games with their products \citep{Duenhaupt2012FinancializationAndRentierIncomeShareUSAGermany}. Many industries tend to get to this point, even if the dominant corporations can be 'creatively destroyed' later for something cheaper, more efficient, or presenting new capability \citep[pp 81-86]{Schumpeter1943CapitalismSocialismAndDemocracy}, rather than simply better marketing.  

\section{Concluding thoughts}  
The $\lambda$ concept that is related to $i$ and $f$ (eq. \ref{eq:SRES_1}) has major application in macroeconomics because it establishes a physics-based metric for evaluating the currency valuation of a nation. I believe this metric can be used to aid management of the money supply and to determine theoretical limits to spending and economic growth. 

This thermodynamic model initiated by \textit{\citeauthor{Garrett2011Arethe} 2011} is important work on some unsolved matters in economics. These ideas are on the leading edge, and there is a need in economics and finance for numerical, physics, and energy grounded approaches such as this thermodynamics work. \textit{\citeauthor{Garrett2022LotkasWheel} 2022} then asks, is there a long arm of history that puts limits on the present? Broadly speaking, I believe the answer is to what degree. The rigid framework based on a constant $\lambda$ is falsified, but there are still limits. The financial growth equation (Appendix \ref{Sect:Growth_equation}) describes this dependence on what came before in the local environment, which conceptually supports a strong role for the past on the  present.  

We still have a long way to go with efficiencies. It is important to recognize that efficiencies are not only technical improvements to engines, as with improved capture of $E_A(t)$ for one specific process, and improving how much of $E_A(t)$ turns into exergy ($E_X(t)$) rather than waste. Efficiencies are also found in material/energy innovation that use 'waste' as product. For instance, the realization that steel slag has chemical composition very close to basalt, basalt is corrosion resistant, and when melted, basalt can be made into products from rock-wool to basaltic rebar for construction, is a business/engineering innovation. Slag is already melted, the energy already expended, thus it makes sense to use it before cooling to make valuable products---energy efficiency of another kind. 

Buckminster Fuller noticed the (fig. \ref{Fig-1_Y_E_Y-E}-D) $\frac{Y}{E}$ trend of 'more with less' in the 1970's, and he asked economists to have a dynamic view of goods and services due to education in STEM \citep[p. 133, 213]{ Fuller1981CriticalPath}, \citep[p. 23-24, 26-27]{Fuller1975Synergetics}. Fuller presented engineering interacting with resources as a flaw in the Malthusian doctrine as defined in his time. This trend of more with less should reach an asymptote based primarily on maximizing $\alpha$ and $\varepsilon$, coefficients of energy efficiency. However, materials, novel engineering, and access to new resources also play a role. 

I think it is a positive outlook for society that net of inflation, real dollar GDP is increasing faster than energy use. A major part of the reason is visible in the Appendix \ref{Appendix:alpha_EfficiencyOfEngines} graph (fig. \ref{Fig_MaxEfficiencyIGS2024_+_Smil_Watts-of-capacity-by-year-and-type}-A) where we see the improvement in human capture of potential energy $E_A(t)$. Exergy ($E_X(t)$) accounting relative to $E_A(t)$ is not widely available at this time, but we can infer improvements because efficiency of electrical motors are high and improving at 70-95\% \citep{Almeida2011StandardsForEfficiencyOfElectricMotors} and much of industrial processes are electricity powered. 

With the improvments discussed here, the thermodynamic model can also deal with fluctuations, and even the fall of civilizations \citep{Bardi2017SenecaEffect}. 

\section{Datasets created for this analysis}
The supplementary data for \textit{\citeauthor{Garrett2022LotkasWheel} 2022}, that also applies to the first paper, is quite complete, with an Excel file containing columns with worked examples.  For the purposes of this examination, there are 8 \underline{new} datasets for this replication validation effort: \textbf{A.} A best effort for gross world product  ($GWP$)  back to the year 1 CE which is the $Y_{Rep}$ dataset (fig. \ref{Fig-1_Y_E_Y-E}-A) for this study.  \textbf{B.} A best effort for exajoules consumption by year going back to year 1, which is $E_{Rep}$ (fig. \ref{Fig-1_Y_E_Y-E}-C) . \textbf{C.} A new $W$ dataset $W_{\sum LW}$ (fig. \ref{Fig-1_Y_E_Y-E}-B) is the result of using equation \ref{eq:W(t)} to perform summation of the $Y_{LW}$ dataset. \textbf{D.} Population estimates back to -14,000 CE $Population$ (fig. \ref{Fig_Morris_W_E_W-E_Pop}-F). \textbf{E.} Exajoules back to -14,000 CE $E_{RepMorris}$ (fig. \ref{Fig_Morris_W_E_W-E_Pop}-C. \textbf{F.} $Y$ back to -14,000 CE $Y_{RepMorris}$ (fig. \ref{Fig_Morris_W_E_W-E_Pop}-A). \textbf{G.} $W$ summed from $Y_{RepMorris}$, $W_{\sum RepMorris}$ (fig. \ref{Fig_Morris_W_E_W-E_Pop}-B). \textbf{H.} $\frac{W}{E}$ including early Morris derived data, $\frac{W_{\sum RepMorris}}{E_{RepMorris}}$ (fig. \ref{Fig_Morris_W_E_W-E_Pop}-E). 

\subsection{$GWP$  interpolated composite dataset---$Y_{Rep}$  }

The primary dataset for $Y_{Rep}$ (the GWP replication dataset) was obtained from \citet[(OWID)]{OWID2024GlobalGDPLongRun}. This dataset has complete sequential data starting 1990. From the year 1 to 1960, intervals of 1000, 500, and 100 years were interpolated using growth function eq. \ref{eq:N_i_eq_N_1.a^x}.
\begin{align}
\begin{split}
 N_i&= N_0 \cdot a^x  \quad \Rightarrow \quad
a= e^{\frac{ln(\frac{N_y}{N_0})}{X_y}} \\
     \text{Where: } i&=0 \leq x \leq y \qquad  N_0 =\text{ the start interpolation range value} \\ 
     N_y&= \text{ the end interpolation range value } \quad X_y= \text{ ending range x value.} 
\label{eq:N_i_eq_N_1.a^x} 
\end{split}
\end{align}

The years from 1820 to 1960, have linear interpolations over the 10 to 30 year intervals. 

For the years 1960 to 1990, the FRED dataset NYGDPMKTPCDWLD \citep{FRED2024NYGDPMKTPCDWLD} was used to estimate GWP. For this period, the OWID dataset had 10 year intervals. Each of the 10 year interval years for which an OWID value existed were scaled against FRED values to generate a scaling multiplier. These multipliers were interpolated linearly over the 10 year interval and used to generate $GWP$  values over the intervals of the OWID data. See figure \ref{Fig-1_Y_E_Y-E}-A. This was required to capture detail present in the \citeauthor{Garrett2022LotkasWheel} 2022 supplement dataset that becomes visible in figure \ref{Fig-1_Y_E_Y-E}-F.

The OWID 2024 $Y_{Rep}$ dataset averages 1.44 times the \textit{Garrett 2022} $GWP$  ($Y_{LW}$), with $\sigma$ = 0.06. (Fig \ref{Fig-1_Y_E_Y-E}-A)  This scaling discrepancy should not have anything read into it, as the $Y_{LW}$ dataset is not an error. The discrepancy is due to a change in how OWID $GWP$  and its data sources are calculated versus other datasets \citep{OWID2024GlobalGDPLongRun}, and the difference is consistent enough.  See fig. \ref{Fig-1_Y_E_Y-E}-A. 

\subsection{$GWP$ Extension of $Y$ back to -14,000 CE derived from Morris using Geary-Khamis concept---$Y_{RepMorris}$}

To generate this dataset, the Year 1 CE ratio of GWP to  $E$ was determined to be $\frac{0.03827 \text{ trillion  1990 \$}}{ exajoule (EJ)}$. This ratio was assumed to be constant going back 14,000 years. The logic of this stems from Morris \citep[p. 91]{Morris2013TheMeasureOfCivilization} where he discusses the total energy capture per-capita per-day and the archaeological record of the material culture. This ratio of production to energy may be higher than it should be because the more primitive the society and its tools, the more energy should be expended on a product basis, but it is the best I can do at this time.

I considered trying to extend this dataset back 50,000 to 150,000 years, but for the current purposes, I do not believe it would make a significant qualitative difference. This dataset covers the appearance of towns at end of the last ice-age, 11,700 years ago \citep{Corrick2020SynchronousClimateLastGlacialPeriod}, \citep[pp. 93-98]{ Morris2013TheMeasureOfCivilization}. See fig. \ref{Fig_Morris_W_E_W-E_Pop}-A.  

If a $Y$ dataset were created going back 150,000 years, per Morris it would raise the $W$ curves in the present by the total of 150,000 yr x 4,000 kCal/cap/day x 4184 joules/kCal x 365.25 days/yr. Assuming 1,000,000 people on average over that time, this should represent about 9.2 trillion Geary-Khamis concept dollars. Human population is not well documented that far back, and is mostly believed to be a few hundred thousand people,  so \$9.2 trillion is probably an outside estimate over that time span. A longer period, or more people in deep time could raise the ratio of $W$ to $E$ in the modern period. Of course, there are arguments for extending this back for 70,000 to 300,000 or more years \citep{Brooks2018Long-distanceStoneTransportPigmentUseInTheEarliestMiddleStoneAge,Henshilwood2011_100000yrOchreWorkshop,Shipton2018_78000yrMiddle-Later-StoneAgeInnovation,Chang2025PaleolithicMigrationKuroshioCurrentRyuku} and perhaps more. Modern humans include Neanderthal DNA as well as Denisovan, probably absorbing Neanderthals in the end \citep{Li2024RecurrentGeneFlowBetweenNeanderthalsAndModernHumansOverPast200kYears}, so arguably these could be included in modern human population numbers, though this would add perhaps 15,000 \citep{Mafessoni2017BetterSupportForSmallPopulationOfNeandertalsAndLongSharedHistoryOfNeandertalsAndDenisovans}. Archaic homo sapiens date to 600,000 years ago. And a deep time cutoff could be the 117,000 years-long near extinction period from -930,000 to -813,000 CE when the human breeding population is estimated at $\approx{}1,280$ \citep{Hu2023GenomicInferenceOfSevereHumanBottleneck800-900Mya}. This period is suggested to be a speciation event. Choosing cutoffs after this deep time speciation bottleneck is somewhat arbitrary, however, even starting with \$10 trillion as the $W(t)$ for the year -14,000 CE as the sum of the previous nearly one million years, this increase to the $W_{RepMorris}$ dataset prior to 1 CE would make minor impact on the present day's ratios (not shown). 

There could be significant improvements to estimates of deep time human economic activity in the future. Work is ongoing since Morris' excellent book was published in 2013, \citep{Freeman2018SynchronizationOfEnergyHolocene,Rathmann2024HumanPopulationDynamicsInUpperPaleolithicEurope,Schmidt2025LargeScaleAndRegionalDemographicResponsesinEuropeFinalPalaeolithic}, and the next 10-20 years may see major work.

\subsection{Two new $W$ datasets: $W_{\sum LW}$, and $W_{RepMorris}$ compared with $W_{LW}$.} 

Two new $W$ datasets were made using the algorithm of equation \ref{eq:W(t)}. The first is a $W_{\sum LW}$, the lower dotted red curve (fig. \ref{Fig-1_Y_E_Y-E}-B), which is the summation of the $Y_{LW}$ supplement dataset (fig. \ref{Fig-1_Y_E_Y-E}-A). The first 1000 years of $W_{\sum LW}$ is thrown away as misleading. The second is $W_{\sum RepMorris}$ (fig. \ref{Fig_Morris_W_E_W-E_Pop}-B)  which is the summation of the $Y_{RepMorris}$ dataset (fig.  \ref{Fig_Morris_W_E_W-E_Pop}-A). 

$W_{LW}$, the supplied dataset, is shown compared to $W_{\sum LW}$ in figure  \ref{Fig-1_Y_E_Y-E}-B. $W_{LW}$ is compared to all of $W_{\sum RepMorris}$ (fig.  \ref{Fig_Morris_W_E_W-E_Pop}-B). Note that the $W_{LW}$ curve in figure \ref{Fig_Morris_W_E_W-E_Pop}-B prior to 1 CE is a projection of a curve fitted to the $W_{LW}$ dataset. This was done for fairness under the assumption that \textit{\citeauthor{Garrett2011Arethe} 2011} would probably have done so with access to the Morris derived dataset. 
This $W_{LW}(t)$ curve equation is: 
\begin{equation}
    W_{LW}(t) = 2.440 \;e^{({5.965\times10^{-4}}\, t)}\quad R^2 = 0.943
\end{equation}
 It is not clear how this $W_{LW}$ dataset was created. It appears that the $W_{LW}$ dataset had to have been made by an equation, and it was clearly not made by summation of the $Y_{LW}$ dataset provided per the specified algorithm (eq. \ref{eq:W(t)}). The Matlab file for the project was not made available. This equation and it's basis would be of interest, because it has correspondence to the Morris dataset built from the ground up. This suggests good intuition. 

\subsection{Exajoules composite dataset---$E_{Rep}$}

The base replicate dataset for $E_{Rep}$ was obtained from Our World in Data (OWID) \citep{Ritchie2020EnergyMix}. OWID based this dataset on a composite of sources. The OWID figures are given in terawatt hours (TWh) separately by energy source. A column was inserted for sum of sources, and a column to convert TWh to exajoules (EJ) using a conversion of 0.0036 EJ per TWh. The OWID dataset has 10 year increments from 1800 to 1960, a 5 year gap from 1960-1965, and then is yearly. From 1800-1965, data was interpolated linearly in the gaps. 

Note that the  \textit{Garrett 2022} $E$ dataset ($E_{LW}$) in 1 CE is 46.17 exajoules, which is 12.8 petawatt-hours---the global energy consumption attained in 1902, which is is clearly an overestimate (fig. \ref{Fig-1_Y_E_Y-E}-C). The year 1 CE population was on the order of 200-300 million, and 1902 CE population was approximately 1.65 billion, with a higher per capita energy use. 

For the year 1 CE, the $E_{Rep}$ exajoules were estimated using Morris \citep[pp. 55, 61, 68, 111,]{Morris2013TheMeasureOfCivilization} data and a population of 226 million. Western, eastern, and Americas civilization populations were assessed at 34, 74, and 6 million respectively. The balance was assigned to hunter-gatherers. Assuming a population of 226 million results in an estimate of 5.45875 EJ in the year 1.

Two interpolation methods were considered.
\textbf{Interpolation method A}. This value of 5.45875 EJ was used as $N_0$ in equation \ref{eq:N_i_eq_N_1.a^x} where $N_{1800}$ (the $N_y$) is 20.3508 \citep{Ritchie2020EnergyMix}. Solving for $a$ yields 1.00073, which is an average annual growth rate of $\approx$0.073\%. See comparative results in figure \ref{Fig-1_Y_E_Y-E}-C. Comparing this curve to the  population curve, by inspection it appears that significant distortion is likely from this interpolation. Interpolation method A was not used.

\textbf{Interpolation method B}. The population curve was used for interpolation from 1 CE to 1800 CE. This curve was selected as superior to method A both by inspection, and because energy should conform well to population during this pre-industrial period. See fig. \ref{Fig-1_Y_E_Y-E}-C.

For discussions below, EJ means exajoule and $E$ is an exajoule dataset. 

\subsection{Extended Exajoules to -14,000 CE composite dataset---$E_{RepMorris}$}
The dataset was extended to -14,000 CE using Morris \citep[pp. 55, 61, 68, 111,]{Morris2013TheMeasureOfCivilization}. Exajoules were calculated using kCal consumption from Morris' tables of kCal/capita/day estimates with population. Population ratios between Western urban society, Eastern urban society, Americas urban society, and hunter-gatherers determined for 1 CE were maintained going back to -14,000 CE to develop global totals. Morris' tables declined to aboriginal baseline even for settlements. 

\subsection{Population dataset}
For the year 1 CE, the \textit{\citeauthor{Garrett2022LotkasWheel} 2022} dataset provides figures of 225.82 million people. The human population in year 1 CE from other sources is a range. OWID provides a figure of 232 million \citep{OWID2023Population}, and census consensus provides a range of 140-400 million, with a mean and median of 270 million \citep{USCensus2022WorldPop}. The \textit{\citeauthor{Garrett2022LotkasWheel} 2022} figures for population are accepted as reasonable. Prior to 1 CE an average of multiple sources was used   \citep{Haub1995HowManyPeople,Goldewijk2011HYDE3.1Human,McEvedy1978,Thomlinson1975,Durand1977HistoricalEstimatesOfWorldPopulation,Gapminder2021}, and values were interpolated linearly.

\backmatter

\bmhead{Supplementary information}
Pre-publication Figshare link. Must be updated with publication DOI after acceptance. \\
https://figshare.com/s/9870ad630c0bbee2e709 \\
Original data is available as link from \textit{\citeauthor{Garrett2022LotkasWheel} 2022}. A copy of this spreadsheet is in the first sheet of the figshare link spreadsheet. Note that the graphs in this spreadsheet were created using Linux LibreOffice Calc 24.8.1.2. Rendering using Microsoft Excel may have artifacts. If there is a question of units the original column can easily be identified by editing the data for a graph. Curve fits are left at defaults in the spreadsheets, thus default precision will be reported by the spreadsheet. Significant digits to report are presented in the paper. Whether digits are reported or not, the spreadsheet maintains its precision to more decimal places internally.




\section*{Declarations}
\begin{itemize}
\item Funding - No funding was received for this work, which began to validate a citation.
\item Competing interests - The author declares no competing interests.
\item Ethics approval and consent to participate - Not applicable.
\item Consent for publication - Not applicable.
\item Data availability - See supplement.
\item Materials availability - Not applicable.
\item Code availability - Not applicable. 
\item Author contribution - This article was authored by 
\end{itemize}

\noindent

\begin{appendices}

\section{Financial growth equation}
\label{Sect:Growth_equation}
\begin{figure}[ht!]
\includegraphics[width=1.0\textwidth]{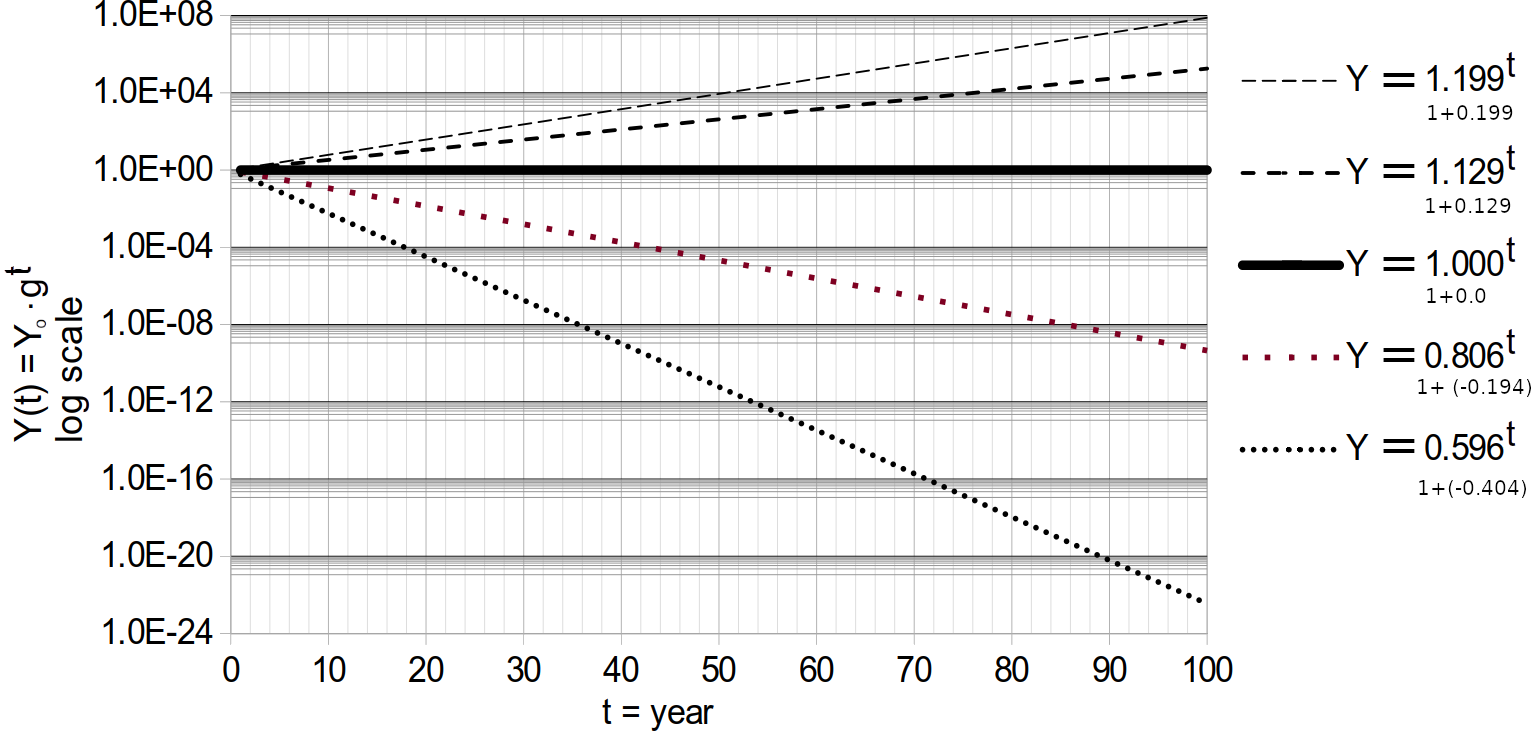}
\caption{Graphs of $Y(t) = Y_0 \cdot g^t$ where $g=1+i$ and $i$ is interest/growth rate. The heavy black curve shows a steady-state zero growth system. In this case, $i=0$ and $g=1$. }
\label{Fig_Y_growth-behavior}
\end{figure}

This growth equation is founded on the empirical fact of capitalism, that the creation of value depends on the capital one has to work with, generally agreeing with \textit{\citeauthor{Garrett2022LotkasWheel} 2022} that the future is dependent on what comes before it. Financiers represent their holdings of capital as units of currency, just as GDP does. 
\begin{align}
\begin{split}
   Y(t) &= Y_0 \cdot g^t \\
   \text{Where: } Y_0 &= \text{ initial value [\$], } g=1+i \text{, and } i = \text{ interest rate}  
   \label{eq:Growth_Yt_Y0xgt}
\end{split}
\end{align}
With this equation, when the economy goes to a steady state, then $g=1$ and $i=0$ (fig. \ref{Fig_Y_growth-behavior}). 

That said, it is clear from fig. \ref{Fig-1_Y_E_Y-E}-D and \ref{Fig_Morris_W_E_W-E_Pop}-D, that for all $\frac{Y}{E}$ curves, the gross world product symbolized by $Y(t)$ has been rising relative to energy since around 1700. So in the real world, if $E(t)$ goes to a steady state of energy consumption, $Y(t)$ should be expected to continue to rise towards an efficiency asymptote relative to $E(t)$, although efficiency has limits set by $\alpha$ and $\varepsilon$. See Appendix \ref{Appendix:alpha_EfficiencyOfEngines}.  

\section{$\frac{dE}{dt}$ and inflation}
\label{Sect:CPI_dE-dt}
The proposition that economic production disappears due to inflation is falsified empirically in two ways.
\begin{quote}
\textbf{Quote \ref{Quote:RealWorldProductionDisappears}}.
    “...real-world economic production disappears: that is, Y = 0. ...note however, that zero real, inflation-adjusted production does not forbid nonzero, positive nominal production. If there is a large difference between the nominal and real GDP, it appears in economic accounts as high values of the GDP deflator or as hyperinflation.” \citep[p.1024]{Garrett2022LotkasWheel} 
 \label{Quote:RealWorldProductionDisappears}    
\end{quote}
\subsection{Empirical test of year on year (YoY) $\frac{dE}{dt}$}
\begin{figure}[!hbt]
\includegraphics[width=0.75\textwidth]{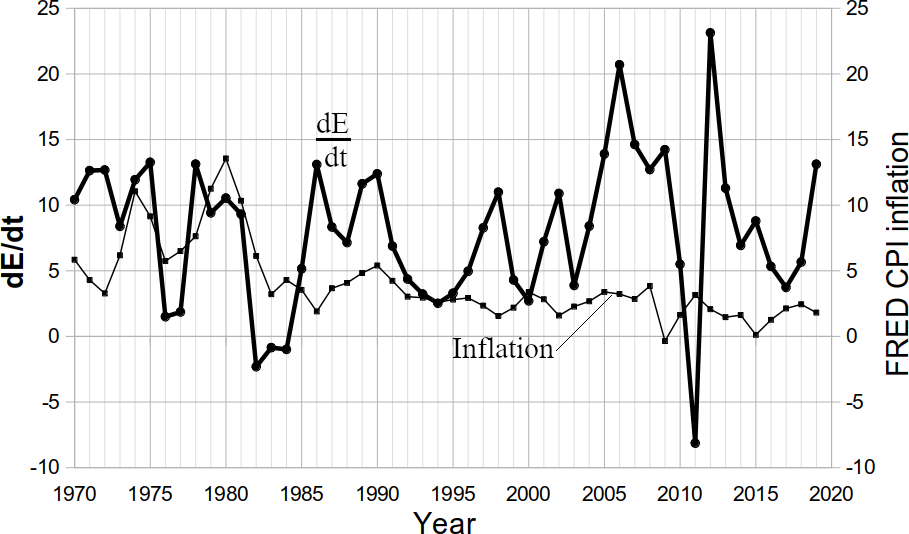}
\caption{YoY $\frac{dE}{dt}$ compared to CPI (inflation). YoY $\frac{dE}{dt}$(heavy black curve) change against FRED CPI inflation (thin black curve) \citep{FRED2024NYGDPMKTPCDWLD}. Here we see that there is no consistent or significant relationship between $\frac{dE}{dt}$ and inflation. There is no suggestion of exponential increase of inflation near zero. When $\frac{dE}{dt}$ crosses zero inflation might rise, or it might fall.}
\label{Fig_dEdt_YoY_to_CPI} 
\end{figure}

Let us examine a comparison (fig. \ref{Fig_dEdt_YoY_to_CPI}) of YoY inflation figures against YoY $\frac{dE}{dt}$ computed as $E_i - E_{i-1}$. If the proposition that $\frac{dE}{dt} = 0$, then $Y = 0$ is true, then it should display behaviour near and at the zero boundary indicating a divide by zero discontinuity. 

The first interesting change in YoY $\frac{dE}{dt}$ in figure \ref{Fig_dEdt_YoY_to_CPI}, is for years 1976-77. These two years, $\frac{dE}{dt}$ dips to 1.5 and 1.9, and inflation drops from 9.1\% in 1975 to 5.7\% in 1976, then rises to 6.5\% in 1977. Next, a dip below zero $\frac{dE}{dt}$ to -2.3, -0.9, and -1.0 in 1982-84 accompanies inflation of 6.1\%, 3.2\%, and 4.3\%, respectively. 2011 has the lowest $\frac{dE}{dt}$ of the dataset at -8.1, and inflation is 3.2\%. The highest inflation year is 1980, at 13.6\%, which is accompanied by $\frac{dE}{dt}$ of 10.5\%. These data do not appear to support the proposition. 

\textit{Garrett 2022} has awareness of the difficulty of resolving 'short-term' behaviours using the model, because $W$ results from a smoothed function \citep[p. 1022]{Garrett2022LotkasWheel}. However, in fig. \ref{Fig_dEdt_YoY_to_CPI}, we are not actually referencing $W$, but instead, we reference $Y$ in context of eq. \ref{eq:Y_estimate_w-hat_eq_Y_over_dE_dt}.  The claim that $Y$ goes to zero when $\frac{dE}{dt}$ is zero, is presented in context of the 50-year \textit{Garrett 2022} study period composed of the discrete data of $E$ and $Y$, which means that those discrete datapoints should conform to the claim. 

One can argue regarding figure \ref{Fig_dEdt_YoY_to_CPI} that a half-century is not a large enough dataset. However, \textit{\citeauthor{Garrett2022LotkasWheel} 2022} is concentrated on this half-century, so if the dataset isn't large enough that would undermine the primary thesis of the paper. One can also argue that $\frac{dE}{dt}$ being zero, near, or below zero is not as prolonged as it would be in a system in balance. However, there is a pair, 1974-75, and a triplet 1980-82, that are close to, and below zero. Those are, therefore, extended periods of 2 and 3 years out of the 50. One can argue that $\frac{dE}{dt}$ has units of exajoules and so even small values of exajoules are still very large, as an exajoule is $10^{18}$ joules. However, because $\frac{dE}{dt}$ is negative for 1980-1982, and 2009, then $\frac{dE}{dt}$ should have passed through zero at least 4 times. This should have exerted effects if the claim were true, which needs to be explained. 

\subsection{Is there an empirical relationship between inflation and $\frac{dE}{dt}$ indicating a discontinuity at zero?}

\begin{figure}[!hbt]
\includegraphics[width=1.0\textwidth]{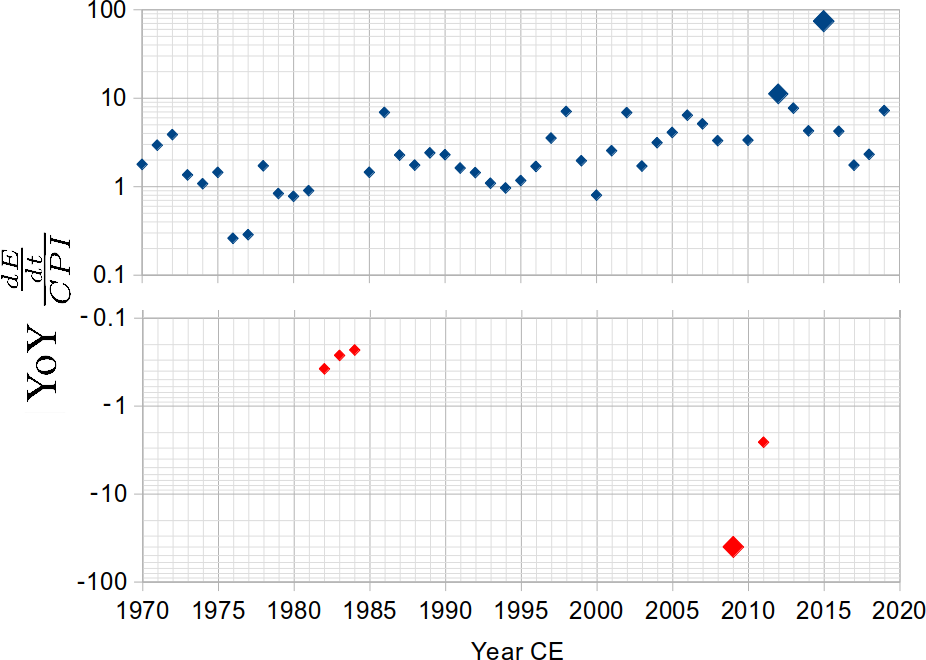}
\caption{Year on Year $\frac{ \frac{dE}{dt} }{CPI}$. Log scale. Note outlier points with an absolute value greater than 10, and negative values. There are no indications of an exponential increase in CPI when $\frac{dE}{dt}$ gets close to, or passes through zero. The relation is mostly flat.} 
\label{Fig_dEdt_div_CPI-neglog}
\end{figure}

Figure \ref{Fig_dEdt_div_CPI-neglog} shows what appears to be a relationship of $\frac{\frac{dE}{dt}}{CPI}$ that appears reasonably stable, save for outliers. Excluding all data with an absolute value less than 10 leaves three outlier points, 2009, 2012, and 2015. A discontinuity at zero is not suggested.

\section{Alpha ($\alpha$) and Epsilon ($\varepsilon$) The efficiency of engines, utilization, and mixtures of them over time}
\label{Appendix:alpha_EfficiencyOfEngines}
\begin{figure}[!hbt]
\includegraphics[width=1.0\textwidth]{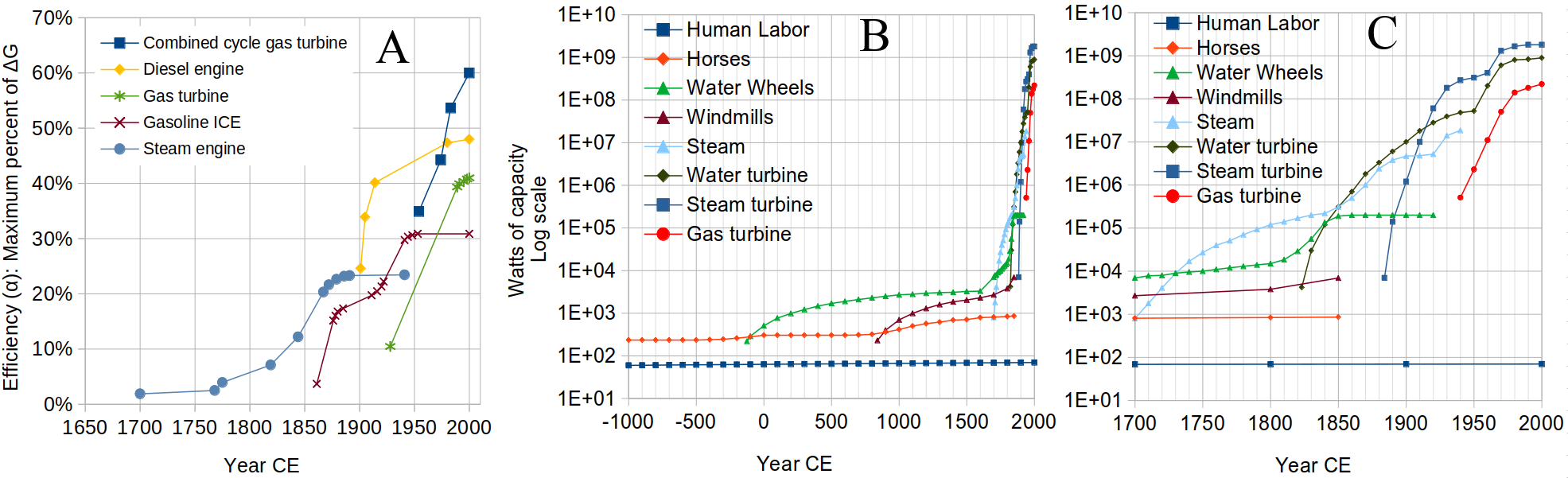}
\caption{\textbf{Panel A.} Efficiency: Maximum percent of $\Delta$G by available energy sources \citep{BostonUniversity_IGS2024MaxEfficienciesOfEnginesAndTurbines1700-2000}. \textbf{Panels B \& C} Watts of capacity, by year and type 1700-2000 \citep{Smil2004WorldHistoryEnergy}. The latter could provide weighted estimates of $\alpha$ by type to develop a rough estimate of civilization's aggregated $\alpha$ over time. }
\label{Fig_MaxEfficiencyIGS2024_+_Smil_Watts-of-capacity-by-year-and-type} 
\end{figure}

$\alpha$ describes the efficiency of engines used to harness energy for human use (fig. \ref{Fig_MaxEfficiencyIGS2024_+_Smil_Watts-of-capacity-by-year-and-type}-A). Note that the efficiencies shown are maximums for top of the line, properly tuned engines. In the real world, each engine of the same type is a bit different, equipment ages, and the mix of power sources can vary a great deal during the course of a day, a year, or from year to year. Figure \ref{Fig_MaxEfficiencyIGS2024_+_Smil_Watts-of-capacity-by-year-and-type}-B \& C shows how watts of energy capacity has varied over time. Each energy category has a different level of maximum efficiency, and the mix supplying energy will vary day by day, even by the second, because electrical grids must balance supply with demand on a second by second basis. In addition, each specific engine type varies from perfect tuning, etcetera. Thus, $\alpha$ can be handled as a function varying in $t$, although $t$ is not causative in the physics sense. 

$\varepsilon \cdot E_A(t)$ gives the fraction, $E_X(t)$, that ends up being used to make products and is not wasted (mostly as heat). $\varepsilon$ should be different for each industrial process in each $\lambda_i$ and over the aggregation of $\Lambda (t)$ should vary in a range from $\approx$70-95\% for electric motors \citep{Almeida2011StandardsForEfficiencyOfElectricMotors}. In the real world, efficiency varies, and economics of electric motor replacement varies as well \citep{GOMEZ2022AssessmentCriteriaOfFeasibilityOfReplacementWithHigh-efficiencyMotors}. Other engines vary a great deal, as shown in figure \ref{Fig_MaxEfficiencyIGS2024_+_Smil_Watts-of-capacity-by-year-and-type}-A. It may be possible to obtain sufficient data to estimate $E_X(t)$ of many industrial processes. 

That said, for most practical purposes $\varepsilon(t)$ will need to be ignored for now because there is little data to work with. $E_A(t)$ is the majority of what is available as energy data. Electricity generation is easy to obtain, as is consumption. For industrial purposes, $E_G(t)$ data would come from consumption of fossil fuel primarily with electrical energy from a variety of sources. $E_A(t)$ is relatively straightforward to obtain, and $E_G(t)$ mostly is a theoretical issue not normally used in economics, except that it is desirable to have the highest possible efficiency conversion of $E_G(t)$ to $E_A(t)$. Over time efforts to track $\varepsilon(t)$ can be made. 

\section{Relation of these equations to Cobb-Douglas}
\label{Appendix:CobbDouglas}
\subsection{Production function, $\Lambda (t)$ and $\lambda_i(t)$ are compatible with the energy based version of Cobb-Douglas (EBCD) }
$\Lambda (t)$ has dimension [$\frac{J}{\$}$]. In this method, the coefficients, $\alpha$ and $\beta$ come from the Cobb-Douglas equation \citep{Cobb-Douglas1928ATheoryOfProduction}. The exponent $\alpha$ is the proportion of earnings accrued to capital. The exponent $\beta$ is the proportion of earnings accrued to labour. Together they multiply to 1, ensuring dimensions remain singular. 

In practice, for each $\lambda_i$, the total capital goods produced and energy consumption are required. These data may be available in reports for many production entities. The exergy of production is unlikely to be available, but the energy consumption of the firms could reasonably be acquired, although this could be a major project. For this reason, I substitute $E_A$ (energy available) for $E_X$ (exergy of productive use). In the case of industrial energy this is mostly electricity consumed after generation. For human energy, it is the available human  energy as measured per Morris \cite[pp. 55, 61, 68, 111]{Morris2013TheMeasureOfCivilization}.

Alternatively, a global function for $\Lambda(t)$ is based on the observation that the units work out so as to be compatible with $\Lambda (t)$. Here, global estimates of each variable in equation \ref{eq:lambda_i_Keen_EBCD}. The degree to which this proposed linkage of the Garrett system of equations to Cobb-Douglas is useful remains to be seen. The system of equations shown in section \ref{Sect:System_of_Equations} can work without it. 

\begin{align}
\begin{split}
 &\text{Cobb-Douglas \citep{Cobb-Douglas1928ATheoryOfProduction}, worked with dimensions and similar substitutions} \\
    \Lambda_i(t) [\frac{J}{\$}] &\equiv(\frac{E_{A_K} [J]}{K [\$]})^\alpha \cdot  (\frac{E_{A_L}[J]}{L [\$]})^{\beta} \\ 
    \text{Where: } \Lambda_i [\frac{J}{\$}]  &= \text{conversion efficiency of Joules into priced products} \\
    \alpha + \beta &= 1 \text{ and } \alpha = \frac{2}{3} \\ 
    L [\$]& = \text{human labor } \quad K \text{ [\$]} = \text{ capital machinery and facilities }\\
    \qquad E_{A_L} \text{ [J]}&= \text{ the  low available energy  of  labor}\\
    E_{A_K} \text{ [J]} &= \text{the high energy consumption of capital}.
    \label{eq:lambda_i_Keen_EBCD}
\end{split}
\end{align} 

\subsection{Energy-Based Cobb-Douglas for conceptual understanding}
This $\Lambda(t) [\frac{J}{\$}]$ aggregate function is where traditional elements that contribute to economics are found. \citeauthor{KEEN2019NoteOnEnergyInProduction} 2019 revised the Cobb-Douglas production function to an Energy Based Cobb-Douglas (EBCD), reproduced here slightly modified, and with dimensions added for ease of understanding. This equation includes $E_{A_K}(t)$ (available energy of capital (K)) and $E_{A_L}(t)$ (available energy of labour) (section \ref{Sect:System_of_Equations}). Thus, $Y(t)$ [\$] is dependent upon $E_A(t)$ (section \ref{Sect:System_of_Equations}) for the respective contributions. Alert readers will notice that as written the EBCD function (eq. \ref{eq:KeenNote1}) produces $Q$ with dimension of [\$$\cdot$J], which conflicts with the usual dimension of GDP\footnote{Per \cite{KEEN2019NoteOnEnergyInProduction} 2019, this is a known issue \citep{Barnett2003DimensionsAndEconomicsSomeProblems,Felipe2020IllusionsOfCalculatingTotalFactorProductivityAndTestingGrowthModels}. It is curable by a $\frac{\$}{J}$ efficiency of $K$ ($\frac{1}{\Lambda(t)}$). I leave this for \citeauthor{KEEN2019NoteOnEnergyInProduction} 2019.}. 
 Substituting modified dimensions for $K$ and $L$ of $\frac{\$}{J}$ fixes the dimensional problem ( eq. \ref{eq:KeenNote2}).
\begin{align}
\begin{split}
\label{eq:KeenNote1}
    &\text{Keen, Energy Based Cobb-Douglas \citep[p 44]{KEEN2019NoteOnEnergyInProduction}}\\
    Q\text{[\$$\cdot$J]} &=(K [\$] \cdot E_{X_K} [J])^\alpha \cdot  (L [\$] \cdot E_{X_L} [J])^{\beta}  \\ 
    \text{Where: } Q \text{[\$$\cdot$J]} &= E \cdot \text{\$ of GDP} \qquad \text{while } \alpha + \beta = 1 \text{ and } \alpha = \frac{2}{3} \\\alpha &= \text{owner profit share; } \beta = \text{labour profit share}  \\
    L [\$] &= \text{human labour} \qquad E_{X_L} \text{ [J]}= \text{low  exergy  of human labour }\\
    K \text{ [\$]}& = \text{capital machinery }\qquad 
    E_{X_K} \text{ [J]} = \text{exergy of capital}
\end{split}
\end{align}

\begin{align}
\begin{split}
\label{eq:KeenNote2}
    &\text{Keen, Energy Based Cobb-Douglas \citep[p 44]{KEEN2019NoteOnEnergyInProduction}}\\
    Q[\$] &=(K [\frac{\$}{J}] \cdot E_{X_K} [J])^\alpha \cdot  (L [\frac{\$}{J}] \cdot E_{X_L} [J])^{\beta}  \\ 
    \text{Where: } Q \text{[\$]} &= \text{\$ of GDP} \qquad \text{while } \alpha + \beta = 1 \text{ and } \alpha = \frac{2}{3} \\\alpha &= \text{owner profit share; } \beta = \text{labour profit share}  \\
    L [\frac{\$}{J}] &= \text{human labour} \qquad E_{X_L} \text{ [J]}= \text{low  exergy  of human labour }\\
    K [\frac{\$}{J}]& = \text{capital machinery }\qquad 
    E_{X_K} \text{ [J]} = \text{exergy of capital}
\end{split}
\end{align}

\end{appendices}
\clearpage

\bibliography{sn-bibliography}

\end{document}